\documentclass[twoside]{article}
\usepackage{fleqn,espcrc2}
\usepackage{graphicx}
\input psfig

\title{Gamma-Ray Bursts, Cosmic-Rays and Neutrinos
\thanks{Invited talk presented at TAUP99, the 6th International Workshop
on Topics in Astroparticle and Underground Physics 
(September 1999, Paris, France).}
}

\author{Eli Waxman
\thanks{Incumbent of the Beracha foundation career development chair}
\thanks{Work supported in part 
by BSF Grant 9800343, AEC Grant 38/99 and MINERVA Grant.}
\address{Dept. of Condensed Matter Physics, Weizmann Institute of Science,
Rehovot 76100, Israel}}

\begin{document}

\begin{abstract}
 
The $\gamma$-ray burst (GRB) model for production of ultra-high-energy, 
$>10^{19}$~eV, cosmic-rays is based on the hypothesis that GRBs arise from 
the dissipation of the kinetic energy of relativistic fireballs at cosmological
distances. Recent observations of delayed low energy emission, ``afterglow,'' 
from GRB sources strongly support the validity of this hypothesis. 
Observations also provide quantitative support for the model.
The inferred physical fireball parameters
imply that protons may be 
accelerated to $>10^{20}$~eV, and the inferred GRB 
energy generation rate is similar to that required to account for the
observed flux of ultra-high-energy cosmic-rays (UHECRs).

Strong suppression of cosmic-ray flux is expected in this model above 
$10^{19.7}$~eV, due to proton interaction with microwave background photons. 
Strong deviations from model flux derived under the assumption of uniform 
source distribution is expected above $10^{20}$~eV, due to source discreteness 
and due to inhomogeneities in source distribution. In particular, the flux 
above $10^{20.5}$~eV is expected to be dominated by few, narrow spectrum 
sources. While model predictions can not be tested (with 
high confidence level) 
using present data, the predicted signatures should be observed with
the planned Auger and Telescope-Array UHECR detectors.

A natural consequence of the GRB model of 
UHECR production is the conversion of 
a large fraction, $\sim10\%$, of the fireball energy to accompanying burst
of $\sim10^{14}{\rm eV}$ and $\sim10^{18}{\rm eV}$ neutrinos. A ${\rm km}^2$ 
neutrino detector would observe several tens of events per year correlated 
with GRBs, and test for neutrino properties (e.g. flavor oscillations,
for which upward moving $\tau$'s would be a unique signature, and coupling
to gravity) with an accuracy many orders 
of magnitude better than is currently possible.

\end{abstract} 

\maketitle

\section{Introduction}

The origin of GRBs,
bursts of 0.1 MeV---1 MeV photons lasting for a few seconds, 
remained unknown for over 20 years, primarily because GRBs 
were not detected prior to 1997 at wave-bands other than $\gamma$-rays 
(see \cite{Fishman} for review of $\gamma$-ray observations).
The isotropic distribution of bursts over the sky
suggested that GRB sources lie at cosmological distances, and
general phenomenological considerations were used to argue that the
bursts are produced by the dissipation of the kinetic
energy of a relativistic expanding fireball (see \cite{fireballs} for
review). 

Adopting the cosmological fireball hypothesis, it
was shown that the physical conditions in the fireball dissipation region 
allow Fermi acceleration of protons
to energy $>10^{20}{\rm eV}$ \cite{W95a,Vietri95}, and that
the average rate at which energy is emitted as $\gamma$-rays
by GRBs is  
comparable to the energy generation rate of UHECRs in a model where
UHECRs are produced by a cosmological distribution of sources \cite{W95b}. 
Based on these two facts, it was suggested that GRBs and UHECRs have
a common origin (see \cite{Nobel_rev} for review). 

In the last two years, afterglows of GRBs have been discovered in X-ray, 
optical, and radio wave bands (see \cite{AG_review} for review).
Afterglow observations confirmed the cosmological origin of the bursts,
through the redshift determination of several GRB host-galaxies
(see \cite{Freedman} for an updated list), and
confirmed \cite{AG_confirm} standard model predictions 
\cite{AG_pred} of afterglows
that result from the collision of an expanding fireball with
its surrounding medium. These observations therefore provide strong
support for the GRB model of UHECR production.

In this review, UHECR and neutrino production in GRBs is discussed in the 
light of recent GRB and UHECR observations. 
The fireball model is briefly described in \S2.1, and proton acceleration
in GRB fireballs is discussed in \S2.2. Recent claims, according to which
protons can not be accelerated to $>10^{20}$~eV in the fireball 
\cite{Gallant98}, are shown in \S2.2 to be erroneous.
Implications of recent afterglow
observations to high energy particle 
production are discussed in \S3. It is 
shown that, contrary to some recent claims \cite{Stecker},
the GRB energy generation rate implied by afterglow observations is 
similar to the energy generation rate required to account for the flux of
$>10^{19}$~eV cosmic-rays.
Model predictions are shown to be consistent with the observed
UHECR spectrum in \S4. 

Predictions
of the GRB model for UHECR production, that can be tested with future 
UHECR experiments, are discussed in \S5. Implications of the
detection by the AGASA experiment of multiple high energy events with
consistent arrival directions \cite{AGASA_pairs} is also discussed in \S5.
High energy neutrino production in fireballs and its implications for future
high energy neutrino detectors are discussed in \S6.

\section{UHECR from GRB fireballs}

\subsection{The fireball model}

In the fireball model of GRBs
\cite{fireball86}, a compact source, of linear scale
$r_0\sim10^7$~cm, produces a wind characterized by an average luminosity 
$L\sim10^{52}{\rm erg\,s}^{-1}$ and mass loss rate $\dot M=L/\eta c^2$.
At small radius, 
the wind bulk Lorentz factor, $\Gamma$, 
grows linearly with radius, until most of the wind energy is converted
to kinetic energy and $\Gamma$ saturates at $\Gamma\sim\eta\sim300$.
Variability of the source on time scale $\Delta t$, resulting
in fluctuations in the wind bulk Lorentz factor $\Gamma$ on similar
time scale, then leads to internal shocks \cite{internal}
in the expanding fireball at a radius
\begin{equation}
r_i\approx\Gamma^2c\Delta t=3\times10^{13}\Gamma^2_{300}\Delta t_{\rm10ms}
{\rm\ cm},
\label{eq:r_i}
\end{equation}
where $\Gamma=300\Gamma_{300}$, $\Delta t=10\Delta t_{\rm10ms}$~ms.
If the Lorentz factor variability within the wind is significant,
internal shocks would reconvert a substantial 
part of the kinetic energy to internal energy. It is assumed that
this energy is then radiated as 
$\gamma$-rays by synchrotron and inverse-Compton emission of
shock-accelerated electrons.

In this model, the observed
$\gamma$-ray variability time, $\sim r_i/\Gamma^2 c\approx\Delta t$,
reflects the variability time of the underlying source, and the GRB
duration, $T\sim10$s, 
reflects the duration over which energy is emitted from the
source. A large fraction of bursts detected by BATSE show variability
on the shortest resolved time scale, $\sim10$~ms \cite{Woods95}, and some show
variability on shorter time scales, $\sim1$~ms \cite{Bhat92}.
This sets the constraint on underlying source size, 
$r_0<c\Delta t\sim10^7$~cm. The wind must be expanding relativistically, 
with a Lorentz factor $\Gamma\sim300$, 
in order that
the fireball pair-production optical depth be small for observed 
high energy, $\sim100$~MeV, GRB photons \cite{Gamma}.

The wind Lorentz factor is expected to fluctuate on
time scales ranging from the source dynamical time, $\Delta t$, to the
wind duration $T$, leading to internal collisions 
over a range of radii, $r\sim r_i=\Gamma^2c\Delta t$ to $r\sim\Gamma^2cT$. 
Internal shocks are generally expected 
to be ``mildly'' relativistic in the fireball 
rest frame, i.e. characterized by Lorentz factor 
$\gamma_i-1\sim1$, since adjacent shells within the wind are expected to
expand with Lorentz factors which do not differ by more than an
order of magnitude. 

As the fireball expands, it drives a relativistic shock (blastwave)
into the surrounding gas. 
At early time, the fireball is little affected by this external interaction. 
At late time, most of the fireball energy is transferred to the 
surrounding gas, and
the flow approaches self-similar expansion.  
For typical fireball parameters, the transition to self-similar expansion
occurs at a radius $r\sim\Gamma^2cT$. At this
radius, mildly relativistic reverse shocks propagate into the fireball
ejecta and decelerate it \cite{reverse}. 
The reverse shocks disappear on (observed)
times scale $T$, and the flow becomes self-similar at later time, with 
a single, relativistic decelerating shock propagating into the surrounding 
medium. Plasma conditions in the reverse shocks are similar to those of 
internal shocks arising from variability on time scale $\sim T$, since both 
are mildly relativistic and occur at similar radii. 
In the discussion that follows
we therefore do not discuss the reverse shocks separately from the 
internal shocks.

The shock driven into the ambient medium continuously heats new gas, and
accelerates relativistic 
electrons that may produce by synchrotron emission 
the delayed radiation, ``afterglow,''
observed on time scales of
days to months.
As the shock-wave decelerates, the emission shifts with time to
lower frequency.

\subsection{Fermi acceleration in GRBs}

In the fireball model, the observed GRB and afterglow radiation is produced
by synchrotron emission of shock accelerated
electrons. In the region where electrons are accelerated, 
protons are also expected to be
shock accelerated. This is similar to what is thought to occur in supernovae 
remnant shocks \cite{Bland87}. We consider below proton acceleration 
in internal (and reverse) fireball shocks.
Since the internal shocks are mildly relativistic,
we expect results related to particle
acceleration in sub-relativistic shocks (see, e.g.,
\cite{Bland87} for review) to be valid for the present
scenario. In particular, the predicted energy distribution of accelerated
protons is $dN_p/dE_p\propto E_p^{-2}$.

Two constraints must be satisfied by
fireball wind parameters in order to allow proton acceleration to
$E_p>10^{20}$~eV in internal shocks \cite{W95a}:
\begin{equation}
\xi_B/\xi_e>0.02\Gamma_{300}^2 E_{p,20}^2L_{\gamma,52}^{-1},
\label{eq:xi_B}
\end{equation}
and
\begin{equation}
\Gamma>130 E_{20}^{3/4}\Delta t^{-1/4}_{10\rm ms}.
\label{eq:G_min}
\end{equation}
Here, $E_p=10^{20}E_{p,20}$~eV, $L_{\gamma}=10^{52}L_{\gamma,52}{\rm erg/s}$
is the $\gamma$-ray luminosity, $\xi_B$ is the 
fraction of the wind energy density which is carried by magnetic field,
$4\pi r^2 c\Gamma^2 (B^2/8\pi)=\xi_B L$, 
and $\xi_e$ is the fraction of wind energy carried by shock
accelerated electrons. Since the electron synchrotron cooling time is short 
compared to the wind expansion time, electrons lose their energy radiatively 
and $L$ is related to the observed  $\gamma$-ray luminosity by
$L_\gamma\approx\xi_e L$. The first condition must be satisfied in order 
for the proton acceleration time $t_a$ to be smaller than the wind expansion 
time. The second condition must be satisfied in order for the 
synchrotron energy loss time of the proton to be larger than $t_a$.

From Eqs. (\ref{eq:xi_B}) and (\ref{eq:G_min}), we infer that 
a dissipative ultra-relativistic wind,
with luminosity and variability time implied by GRB observations,
satisfies the constraints necessary to allow the acceleration of protons 
to energy $>10^{20}$~eV, provided that the wind bulk Lorentz factor is
large enough, $\Gamma>100$, and that the
magnetic field is close to equipartition with electrons. The former 
condition, $\Gamma>100$, is remarkably similar to that inferred based on
$\gamma$-ray spectra. There is no theory at present that allows a basic
principles calculation of the strength of the magnetic field. However, 
magnetic field close to equipartition, $\xi_B\sim1$, is required
in order to account
for the observed $\gamma$-ray emission (see also \S3). 

We have assumed in the discussion so far that the fireball is spherically 
symmetric. However, since a jet-like fireball behaves as if it were
a conical section of a spherical fireball as long as the jet opening
angle is larger than $\Gamma^{-1}$, our
results apply also for a jet-like fireball 
(we are interested only in processes that occur when
the wind is ultra-relativistic, $\Gamma\sim300$, prior to 
significant fireball deceleration). For a jet-like wind, $L$ in our
equations should be understood as the luminosity the fireball
would have carried had it been spherically symmetric.

It has recently been pointed out in \cite{Gallant98} that
conditions at the {\it external, highly relativistic}
shock driven by the fireball into the ambient
gas are not likely to allow proton acceleration to ultra-high energy. 
Although correct, this observation
is irrelevant to the scenario considered here based on \cite{W95a}, 
since in this scenario protons are accelerated in {\it internal, mildly
relativistic} fireball shocks.

\section{Implications of afterglow observations}

In addition to providing support to the validity of the qualitative fireball 
scenario described in \S2.1 \cite{AG_review}, afterglow observations provide 
quantitative constraints on fireball model parameters.

The determination of GRB redshifts implies that
the characteristic GRB $\gamma$-ray luminosity and emitted energy, 
in the 0.05 to 2~MeV band, are $L_\gamma\sim10^{52}{\rm erg/s}$ and 
$E_\gamma\sim10^{53}{\rm erg}$ respectively (e.g. \cite{Freedman}), 
an order of magnitude higher than the values
assumed prior to afterglow detection
(here, and throughout the paper, we assume an open universe, 
$\Omega=0.2$, $\Lambda=0$, and $H_0=75{\rm\ km/s\ Mpc}$). 
The increased GRB luminosity scale implies that the constraint 
(\ref{eq:xi_B}) on the fireball magnetic field  
is less stringent than previously assumed. 

Due to present technical limitations of the experiments, 
afterglow radiation is observed in most cases only on time scale $>>10$~s.
At this stage, radiation is produced by the
external shock driven into the surrounding gas, and afterglow observations
therefore do not provide direct constraints on the magnetic field 
energy fraction $\xi_B$ at the internal and reverse shocks, where
protons are accelerated to ultra-high energy. In one case, however, 
that of GRB~990123, reverse shock emission was detected over $\sim10$~s
time scale \cite{Akerlof99,SPMR_0123}. 
For this case, the inferred value of $\xi_B$ \cite{WnD00} is consistent with
the constraint (\ref{eq:xi_B}). Clearly, more observations are required
to determine whether this condition is generally satisfied.

The observed GRB redshift distribution implies a GRB rate of 
$R_{\rm GRB}\sim10/{\rm Gpc}^3{\rm yr}$ at $z\sim1$. 
The present, $z=0$, rate is less well constrained, since most observed 
GRBs originate at redshifts $1\le z\le2.5$ \cite{GRB_z}. 
Present data
are consistent with both no evolution of GRB rate with redshift, and 
with strong evolution (following, e.g.,
the luminosity density evolution of QSOs or the evolution of
star formation rate), in which $R_{\rm GRB}(z=1)/R_{\rm GRB}(z=0)\sim8$
\cite{GRB_z}.
The energy observed in $\gamma$-rays reflect the fireball
energy in accelerated electrons. Afterglow observations imply that 
accelerated electrons and protons carry similar energy \cite{Freedman}.
Thus, the inferred $z=0$ rate of cosmic-ray production by GRBs is
similar to the generation rate of $\gamma$-ray energy, 
\begin{equation}
E^2 (d\dot n_{CR}/dE)_{z=0}= 10^{44}\zeta {\rm erg/Mpc}^3{\rm yr},
\label{eq:cr_rate}
\end{equation}
where $\zeta$ is in the range of $\sim1$ to $\sim8$.
This energy generation rate is remarkably similar to that 
implied by the observed UHECR flux (see \S4).\footnote{
It has recently been argued \cite{Stecker} that
the $z=0$ GRB $\gamma$-ray energy generation rate
is much smaller, $\sim10^{42}{\rm erg/Mpc}^3{\rm yr}$. 
Most of the discrepancy between this result and our result can be accounted 
for by noting two errors made in the analysis of ref. \cite{Stecker}: 
estimating the energy generation rate as the product of the GRB rate and 
the {\it median}, rather than {\it average}, GRB energy, and using
(following \cite{MnM98}) the GRB
energy observed in the 50 to 300~keV band, where only a small fraction 
of the 0.05 to 2~MeV $\gamma$-ray energy is observed.}

\section{Comparison with UHECR observations}

Fly's Eye \cite{Bird934} and AGASA \cite{Hayashida945,Takeda98} results
confirm the flattening of the cosmic-ray spectrum at $\sim10^{19}$~eV,
\begin{figure}
\centerline{\psfig{figure=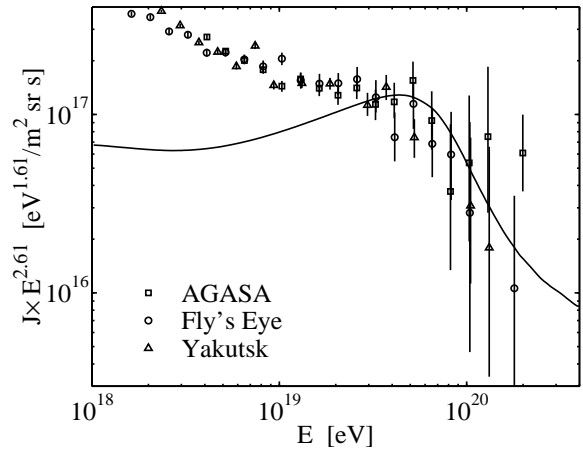,width=3in}}
\caption{
The UHECR flux expected in a cosmological model, where high-energy protons 
are produced at a rate $(E^2 d\dot n_{CR}/dE)_{z=0}=0.8\times10^{44}
{\rm erg/Mpc}^3{\rm yr}$ as predicted in the GRB model 
[Eq. (\ref{eq:cr_rate})], compared to the Fly's Eye, Yakutsk and AGASA data. 
$1\sigma$ flux error bars are shown. The highest energy points are derived
assuming the detected events (1 for Fly's Eye and Yakutsk, 
4 for AGASA) represent a
uniform flux over the energy range $10^{20}$~eV--$3\times10^{20}$~eV.}
\label{fig1}
\end{figure}
evidence for which existed in previous experiments with weaker statistics
\cite{Watson91}. Fly's Eye data is well fitted in the energy range 
$10^{17.6}$~eV to $10^{19.6}$~eV by a sum of two power laws: A
steeper component, with differential number spectrum
$J\propto E^{-3.50}$, dominating at lower
energy, and a shallower component, $J\propto E^{-2.61}$, 
dominating at higher energy, $E>10^{19}$~eV.
The flattening of the spectrum, combined with the lack of anisotropy 
and the evidence for a change in composition from heavy nuclei at low
energy to light nuclei (protons) at high energy \cite{composition},
suggest that an extra-Galactic source of protons dominates the flux at
high energy $E>10^{19}$~eV.

In Fig. 1 we compare the UHECR spectrum,
reported by the Fly's Eye \cite{Bird934}, the Yakutsk \cite{Yakutsk}, 
and the AGASA \cite{Takeda98} experiments, 
with that predicted by the GRB model.
The proton generation rate is assumed to evolve in redshift 
following QSO luminosity evolution \cite{QSO}.
Note, that the cosmic-ray spectrum at energy $>10^{19}$~eV is  
little affected by modifications of the cosmological parameters or of
the redshift evolution of cosmic-ray generation rate, since
cosmic-rays at this energy originate from distances shorter than
several hundred Mpc. The spectrum and flux at $E>10^{19}$~eV is mainly 
determined by the present ($z=0$) generation rate and spectrum.
The absolute flux measured at
$3\times10^{18}$~eV differs between the various experiments,
corresponding to a systematic $\simeq10\%$ ($\simeq20\%$) over-estimate 
of event energies in the AGASA (Yakutsk)
experiment compared to the Fly's Eye experiment (see also \cite{Hayashida945}).
In Fig. 1, the Yakutsk energy normalization is used. 

The suppression of model flux above $10^{19.7}$~eV is  
due to energy loss of high energy protons
in interaction with the microwave background, i.e. to the ``GZK cutoff''
\cite{GZK}. 
Both Fly's Eye and Yakutsk data show a deficit in the number of events,
compared to the number expected based on extrapolation of the 
$J\propto E^{-2.61}$ power-law fit, consistent with the predicted 
suppression. The deficit is, however, only
at a $2\sigma$ confidence level \cite{W95a}.
The AGASA data is consistent 
with Fly's Eye and Yakutsk results below $10^{20}$~eV.
A discrepancy may be emerging at higher energy, $>10^{20}$~eV, 
where the Fly's Eye and Yakutsk experiments detect 1 event each,
and the AGASA experiment detects 6 events for similar exposure. 

The flux above $10^{20}{\rm eV}$ is
dominated by sources at distances $<30\ {\rm Mpc}$ \cite{W95b} 
(see \S5). Since the distribution of known astrophysical systems
(e.g. galaxies, clusters of galaxies) is inhomogeneous on scales of
tens of Mpc, significant deviations from model predictions presented
in Fig. 1 for a uniform source distribution are expected above 
$10^{20}{\rm eV}$. It has recently been shown \cite{CR_clustering}
that clustering of cosmic-ray sources leads
to a standard deviation, $\sigma$, in the expected number, $N$, of 
events above $10^{20}$ eV, given by 
$\sigma /N = 0.9(d_0/10 {\rm Mpc})^{0.9}$, where $d_0$ is the unknown scale
length of the source correlation function and $d_0\sim10$ Mpc 
for field galaxies.

An order of magnitude increase in the exposure of UHECR experiments,
compared to that available at present, is required to test for
the existence of the GZK
cutoff \cite{W95b}. Such exposure would allow this test through an  
accurate determination of the spectrum in the energy range
of $10^{19.7}$~eV to $10^{20}$~eV, where the effects of source inhomogeneities
are expected to be small \cite{W95b,CR_clustering}. 
Moreover, an order of magnitude
increase in exposure will also allow to determine the source correlation 
length $d_0$, through the detection of anisotropies in the arrival directions
of $\sim10^{19.5}$~eV cosmic-rays over angular scales of 
$\Theta\sim d_0/30$~Mpc \cite{CR_clustering}.

Finally, we note that preliminary results from the HiRes experiment 
were presented in this conference \cite{Matthews99}, reporting 7 events beyond 
$10^{20}$~eV for an exposure similar to that of the Fly's Eye. It is difficult
to decide how to interpret this result, since the discrepancy between 
HiRes and Fly's Eye results is present not only above $10^{20}$~eV but also 
at lower energy, where Fly's Eye, AGASA and Yakutsk experiments are in 
agreement: 13 events above $6\times10^{19}$~eV are reported in the preliminary
HiRes analysis, while only 5 events at that energy range are reported by the
Fly's Eye. We therefore believe that unambiguous conclusions based on the 
recent HiRes data can only be drawn after a complete analysis of the HiRes 
data is published.

\section{GRB model predictions for planned UHECR experiments}

The energy of the most
energetic cosmic ray detected by the Fly's Eye experiment is in excess of
$2\times10^{20}{\rm eV}$, and that of the most
energetic AGASA event is $\sim2\times10^{20}{\rm eV}$. On a
cosmological scale, the distance traveled by such energetic particles is
small: $<100{\rm Mpc}$ ($50{\rm Mpc}$) for the AGASA (Fly's Eye) event
(e.g., \cite{Aharonian94}). Thus, the detection of these events over a $\sim5
{\rm yr}$ period can be reconciled with the rate of nearby GRBs, $\sim1$
per $100\, {\rm yr}$ out to $100{\rm Mpc}$, only if
there is a large dispersion, $\geq100{\rm yr}$, in the arrival time of protons 
produced in a single burst. 

The required dispersion
is likely to occur due to the combined effects of deflection 
by random magnetic fields and energy dispersion of the particles
\cite{W95a}. 
A proton of energy $E$ propagating over a distance $D$
through a magnetic field of strength $B$ and correlation length
$\lambda$ is deflected by an angle
$\theta_s\sim(D/\lambda)^{1/2}\lambda/R_L$, which results in a time
delay, compared to propagation along a straight line,
$\tau(E,D)\approx\theta_s^2D/4c\propto B^2\lambda$. 
The random energy loss UHECRs suffer as they propagate, owing to the 
production of pions, implies that 
at any distance from the observer there is some finite spread
in the energies of UHECRs that are observed with a given fixed energy.
For protons with energies $>10^{20}{\rm eV}$ 
the fractional RMS energy spread is of order unity
over propagation distances in the range $10-100{\rm Mpc}$ 
(e.g. \cite{Aharonian94}).
Since the time delay is sensitive to the particle energy, this implies that
the spread in arrival time of UHECRs with given observed energy is comparable
to the average time delay at that energy, $\tau(E,D)$.

The magnetic field required in order to produce a spread 
$\tau(E=10^{20}{\rm eV},D=100{\rm Mpc})>100$~yr, is 
well below the current upper bound on the inter-galactic magnetic 
field, $B\lambda^{1/2}\le10^{-9}{\rm G\ Mpc}^{1/2}$ \cite{IGM},
which allows a spread $\tau\sim10^5$~yr. 
We discuss below some implications, unique to the GRB model, of 
time delays induced by magnetic-field deflection.

\subsection{The highest energy sources}

The rapid increase with energy of the pion production energy loss rate
effectively
introduces a cutoff distance, $D_c(E)$, beyond which sources do not contribute
to the flux above $E$. The function $D_c(E)$ is shown in Fig. 2. 
We define a critical energy $E_c$, for which the average number of sources
at $D<D_c(E_c)$ is 1, 
$\frac{4\pi}{5} R_{GRB}D_c(E_c)^3 \tau\left[E_c,D_c(E_c)\right]=1$ 
\cite{MnW96}.
Although $E_c$ depends through $\tau$ on the unknown properties of 
the intergalactic magnetic field, the rapid
decrease of $D_c(E)$ with energy near $10^{20}{\rm eV}$
implies that $E_c$ is only weakly dependent on the value of $B^2\lambda$. 
In The GRB model, the product $R_{GRB}\tau(D=100{\rm Mpc},E=10^{20}{\rm eV})$
is approximately limited to the range $10^{-6}{\rm\ Mpc}^{-3}$ to
$10^{-3}{\rm\ Mpc}^{-3}$ (The lower limit is set by the requirement that 
at least a few GRB sources be present at $D<100$~Mpc, and the upper limit by 
the Faraday rotation bound 
$B\lambda^{1/2}\le10^{-9}{\rm G\ Mpc}^{1/2}$ \cite{IGM} and 
$R_{GRB}\le10/{\rm\ Gpc}^3{\rm yr}$). The corresponding range
of values of $E_c$ is 
$10^{20}{\rm eV}\le E_c<3\times10^{20}{\rm eV}$.

Fig. 2 presents the flux obtained in one realization of
a Monte-Carlo simulation described in ref.
\cite{MnW96} of the total
\begin{figure}
\centerline{\psfig{figure=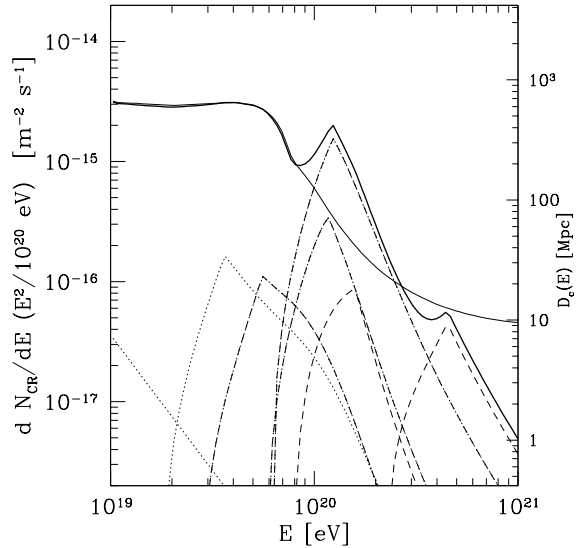,width=3in}}
\caption{Results of a Monte-Carlo realization of the bursting sources
model, with $E_c=1.4\times10^{20}$~eV: Thick solid line- overall 
spectrum in the realization;
Thin solid line- average spectrum, this
curve also gives $D_c(E)$;
Dotted lines- spectra of brightest sources at different energies.
}
\label{figNc}
\end{figure}
number of UHECRs received from GRBs at some fixed time for 
$E_c=1.4\times10^{20}$~eV. 
For each
realization the distances and
times at which cosmological GRBs occurred were randomly drawn. 
Most of the realizations gave an overall spectrum similar to that presented
in Fig. \ref{figNc} when the brightest source of this 
realization (dominating at $10^{20}{\rm eV}$) is not included.
At $E < E_c$,
the number of sources contributing to the flux is very large, 
and the overall UHECR flux received at any
given time is near the average flux (obtained for spatially and temporally
homogeneous UHECR volume emissivity).
At $E > E_c$, the flux will generally be much lower than the average,
because there will be no burst within a distance $D_c(E)$ having taken
place sufficiently recently. There is, however, a significant probability
to observe one source with a flux higher than the average.
A source similar to the brightest one in Fig. \ref{figNc}
appears $\sim5\%$ of the time. 

At any fixed time a given burst is observed in UHECRs only over a narrow
range of energy, because if
a burst is currently observed at some energy $E$ then UHECRs of much lower
(higher) energy from this burst will arrive (have arrived) mainly in the 
future (past). For energies above the 
pion production threshold, 
$E>10^{19.7}{\rm eV}$, the dispersion in arrival times of UHECRs
with fixed observed energy is comparable to the average delay at that
energy. This implies that
the spectral width $\Delta E$ of the source at a given time is of order
the average observed energy, $\Delta E\sim E$.
Thus, bursting UHECR sources should have narrowly peaked energy
spectra,
and the brightest sources should be different at different energies.
For steady state sources, on the other hand, the brightest
source at high energies should also be the brightest one at low
energies, its fractional contribution to the overall flux decreasing to
low energy only as $D_c(E)^{-1}$.
A detailed numerical analysis of the time dependent energy spectrum of 
bursting sources is given in \cite{Sigl_Lemoine97}.

The AGASA experiment reported the presence
of one triplet and three doublets of UHECRs with angular separations 
(within each multiplet) 
$\le2.5^\circ$, roughly consistent with the measurement error,
among a total of 47 UHECRs with $E\ge4\times10^{19}{\rm eV}$ 
\cite{AGASA_pairs}. The probability to have found such multiplets by chance
is $\sim1\%$. Therefore, this observation favors
the bursting source model, although more data are needed to confirm it.

Testing the GRB model predictions described above 
requires an exposure 10 times larger than that of present
experiments. Such increase is expected to be provided by the planned
Auger \cite{Auger} and Telescope Array \cite{TA} detectors.

\subsection{Spectra of Sources at $E<4\times10^{19}{\rm eV}$}
\label{subsec:Blambda}

For nearby, $D<100$~Mpc, sources contributing at $E\le4\times10^{19}{\rm eV}$,
pion production energy loss is negligible, and particle energy may be 
considered constant along the propagation path. 
In this case, the spectral shape of individual sources depends primarily on
the magnetic field correlation length \cite{WnM96}. 

If $\lambda \gg D\theta_s(D,E)
\simeq D(D/\lambda)^{1/2}\lambda/R_L$, all UHECRs that arrive at the
observer are essentially deflected by the same magnetic field structures, 
and the absence of random energy loss during propagation implies that
all rays with a fixed observed energy would reach the observer with exactly
the same direction and time delay. At a fixed time, therefore, the source would
appear mono-energetic and point-like (In reality,
energy loss due to pair production
results in a finite but small spectral and angular width, 
$\Delta E/E\sim\delta\theta/\theta_s\le1\%$ \cite{WnM96}).

If, on the other hand, $\lambda \ll D\theta_s(D,E)$,
the deflection of different UHECRs arriving at the observer
are essentially independent. Even in the absence of any energy loss there 
are many paths from the source to the observer for UHECRs of fixed energy $E$
that are emitted from the source at an angle 
$\theta\le\theta_s$ relative to the source-observer line of sight. Along
each of the paths, UHECRs are deflected by independent magnetic field 
structures. Thus, the source angular size would be of order $\theta_s$
and the spread in arrival times would be comparable to the characteristic 
delay $\tau$, leading to $\Delta E/E\sim1$ 
(The spectral shape of sources is given
in analytic form for this case in \cite{WnM96}).

For $D=30{\rm Mpc}$ and $E\simeq10^{19}{\rm eV}$, 
the $\theta_s D=\lambda$ line divides the
allowed region (for the GRB model) 
in the $B$--$\lambda$ plane at $\lambda\sim1{\rm Mpc}$. Thus, 
measuring the spectral width of bright sources would allow to determine
if the field correlation length is much larger, much smaller, or comparable
to $1{\rm Mpc}$.

\section{High energy Neutrinos}

\subsection{GRB neutrinos, $\sim10^{14}$~eV}

Protons accelerated in the fireball to high energy lose energy through
photo-meson interaction with fireball photons. The decay of charged
pions produced in this interaction, $\pi^+\rightarrow\mu^++\nu_\mu
\rightarrow e^++\nu_e+\overline\nu_\mu+\nu_\mu$, 
results in the production of high energy neutrinos \cite{WnB97}. 
The neutrino spectrum is determined by the observed gamma-ray
spectrum, which is well described by a broken power-law,
$dN_\gamma/dE_\gamma\propto E_\gamma^{-\beta}$ 
with different values of $\beta$ at low and high energy \cite{Fishman}. The
observed break energy (where $\beta$ changes) is typically 
$E_\gamma^b\sim1{\rm MeV}$, 
with $\beta\simeq1$ at energies below the break and $\beta\simeq2$ 
above the break. The interaction of protons accelerated to a power-law
distribution, $dN_p/dE_p\propto E_p^{-2}$, 
with GRB photons results in a broken power law
neutrino spectrum, $dN_\nu/dE_\nu\propto E_\nu^{-\beta}$ with
$\beta=1$ for $E_\nu<E_\nu^b$, and $\beta=2$ for $E_\nu>E_\nu^b$. 
The neutrino break energy $E_\nu^b$ is fixed by the threshold energy
of protons for photo-production in interaction with the dominant $\sim1$~MeV
photons in the GRB \cite{WnB97},
\begin{equation}
E_\nu^b\approx5\times10^{14}\Gamma_{300}^2(E_\gamma^b/1{\rm MeV})^{-1}{\rm eV}.
\label{Enu}
\end{equation}

The normalization of the flux is determined by the
efficiency of pion production.
As shown in \cite{WnB97}, the fraction of energy lost to pion production
by protons producing the neutrino flux above the break, $E^b_\nu$, is 
essentially independent of energy and is given by
\begin{equation}
f_\pi\approx0.2{L_{\gamma,52}\over
(E_\gamma^b/1{\rm MeV})\Gamma_{300}^4 \Delta t_{10\rm  ms}}.
\label{fpi}
\end{equation}
Thus, acceleration of protons to high energy in internal fireball
shocks would lead to conversion of a significant fraction of
proton energy to high energy neutrinos.

If GRBs are the sources of UHECRS, 
then using Eq. (\ref{fpi}) and the UHECR generation rate
given by Eq. (\ref{eq:cr_rate}) with $\zeta\simeq1$,
the expected GRB neutrino flux is \cite{WnB99}
\begin{eqnarray}
E_\nu^2\Phi_{\nu_x}\approx&&
1.5\times10^{-9}\left({f_\pi\over0.2}\right)\times\cr
&&\min\{1,E_\nu/E^b_\nu\}
{{\rm GeV}\over{\rm cm}^{2}{\rm s\,sr}},
\label{JGRB}
\end{eqnarray}
where $\nu_x$ stands for $\nu_\mu$, $\bar\nu_\mu$ and $\nu_e$.

The neutrino spectrum (\ref{JGRB}) is
modified at high energy, where neutrinos are produced by the decay
of muons and pions whose life time 
exceeds the characteristic time for
energy loss due to adiabatic expansion and synchrotron emission 
\cite{WnB97,RnM98,WnB99}.
The synchrotron loss time is determined by the energy density of the
magnetic field in the wind rest frame.
For the characteristic parameters of a GRB wind, 
synchrotron losses are the dominant effect, leading to strong suppression of
$\nu$ flux above
$\sim10^{16}$~eV.

We note, that the results presented above were derived using the 
``$\Delta$-approximation,'' i.e.
assuming that photo-meson interactions are dominated by the contribution of
the $\Delta$-resonance.
It has recently been shown \cite{Muecke98}, that for photon spectra harder
than $dN_\gamma/dE_\gamma\propto E^{-2}_\gamma$, the contribution of 
non-resonant interactions may be important. Since in order to interact with
the hard part of the photon spectrum, $E_\gamma<E_\gamma^b$, the proton energy
must exceed the energy at which neutrinos of energy $E_\nu^b$ are
produced, significant modification of the $\Delta$-approximation results
is expected only for $E_\nu\gg E_\nu^b$, where the neutrino flux is 
strongly suppressed by synchrotron losses.

\subsection{Afterglow neutrinos, $\sim10^{18}$~eV}

Protons are expected to be accelerated to 
$>10^{20}$~eV in both internal shocks due to variability of the underlying
source, and in the reverse shocks driven into the fireball ejecta at the
initial stage of interaction of the fireball with its surrounding gas, which
occurs on time scale $T\sim10$~s, comparable to the duration of the GRB itself.
Optical--UV photons are radiated  by electrons accelerated in shocks
propagating backward into the ejecta, and may interact with accelerated
protons. The interaction of these low energy, 10~eV--1~keV, photons 
and high energy protons produces a burst of duration $\sim T$ 
of ultra-high energy, $10^{17}$--$10^{19}$~eV, neutrinos [as indicated by
Eq. (\ref{Enu})] via photo-meson interactions \cite{AG_nus}.

Afterglows have been detected in several cases; reverse
shock emission has only been identified for GRB 990123 \cite{Akerlof99}.
Both the detections and the non-detections are consistent with shocks
occurring with typical model parameters \cite{SPMR_0123},
suggesting that reverse shock emission may be common.
The predicted neutrino emission depends, however, upon parameters
of the surrounding medium that
can only be estimated once
more observations of the prompt optical afterglow emission are available.

If the density of gas surrounding the fireball is 
typically $n\sim1{\rm cm}^{-3}$,
a value typical to the inter-stellar medium and consistent with
GRB 990123 observations, then the expected neutrino intensity is
\cite{AG_nus}
\begin{equation}
E_\nu^2\Phi_{\nu_x}\approx 10^{-10}
\left({E_\nu\over10^{17}{\rm eV}}\right)^{\beta}
{{\rm GeV}\over{\rm cm}^{2}{\rm s\, sr}},
\label{eq:Phinu}
\end{equation}
where $\beta=1/2$ for $\epsilon_\nu^{\rm ob.}>10^{17}{\rm eV}$
and $\beta=1$ for $\epsilon_\nu^{\rm ob.}<10^{17}{\rm eV}$.
Here too, $\nu_x$ stands for $\nu_\mu$, $\bar\nu_\mu$ and $\nu_e$.
The neutrino flux is expected to be strongly suppressed at energy
$>10^{19}$~eV, since protons are not expected to be
accelerated to energy $\gg10^{20}$~eV. 

The neutrino flux due to interaction with reverse shock photons may be 
significantly higher than that given in Eq. (\ref{eq:Phinu}), if the
density of gas surrounding the fireball is significantly higher than 
the value we have assumed, i.e. if $n\gg1{\rm cm}^{-3}$.

\subsection{Implications}

The flux of $\sim10^{14}$~eV neutrinos given in Eq. (\ref{JGRB}) implies
that large area, $\sim1{\rm km}^2$, high-energy neutrino telescopes,
which are being constructed to detect 
cosmologically distant neutrino sources (see \cite{Halzen_review99}
for review), would
observe several tens of events per year correlated with GRBs. 
The detection rate of ultra-high energy, $\sim10^{18}$~eV, afterglow 
neutrinos implied by Eq. (\ref{eq:Phinu}) is much lower. The
$\sim10^{18}$~eV neutrino flux depends,
however, on parameters of the surrounding medium which can be estimated
only once more observations of reverse shock emission are available.

One may look
for neutrino events in angular coincidence, on degree scale, 
and temporal coincidence, on time scale of seconds, with GRBs \cite{WnB97}. 
Detection of neutrinos from GRBs could be used to
test the simultaneity of
neutrino and photon arrival to an accuracy of $\sim1{\rm\ s}$
($\sim1{\rm\ ms}$ for short bursts), checking the assumption of 
special relativity
that photons and neutrinos have the same limiting speed.
These observations would also test the weak
equivalence principle, according to which photons and neutrinos should
suffer the same time delay as they pass through a gravitational potential.
With $1{\rm\ s}$ accuracy, a burst at $100{\rm\ Mpc}$ would reveal
a fractional difference in limiting speed 
of $10^{-16}$, and a fractional difference in gravitational time delay 
of order $10^{-6}$ (considering the Galactic potential alone).
Previous applications of these ideas to supernova 1987A 
(see \cite{John_book} for review), where simultaneity could be checked
only to an accuracy of order several hours, yielded much weaker upper
limits: of order $10^{-8}$ and $10^{-2}$ for fractional differences in the 
limiting speed and time delay respectively.

The model discussed above predicts the production of high energy
muon and electron neutrinos. 
However, if the atmospheric neutrino anomaly has the explanation it is
usually given, oscillation to $\nu_\tau$'s with mass $\sim0.1{\rm\ eV}$
\cite{atmo}, then
one should detect equal numbers of $\nu_\mu$'s and $\nu_\tau$'s. 
Up-going $\tau$'s, rather than $\mu$'s, would be a
distinctive signature of such oscillations. 
Since $\nu_\tau$'s are not expected to be produced in the fireball, looking
for $\tau$'s would be an ``appearance experiment.''
To allow flavor change, the difference in squared neutrino masses, 
$\Delta m^2$, should exceed a minimum value
proportional to the ratio of source
distance and neutrino energy \cite{John_book}. A burst at $100{\rm\ Mpc}$ 
producing $10^{14}{\rm eV}$ neutrinos can test for $\Delta m^2\ge10^{-16}
{\rm eV}^2$, 5 orders of magnitude more sensitive than solar neutrinos.

\end{document}